\documentclass[pra,aps,groupaddress,twocolumn, nofootinbib,longbibliography]{revtex4-2}
\usepackage[margin=1in]{geometry}

\usepackage{amsmath,amssymb,mathtools,xcolor}
\usepackage{bm}
\usepackage{diagbox}
\usepackage{hyperref}
\usepackage{tikz,pgfplots,pgfplotstable}
\usepackage{changes}

\usetikzlibrary{positioning}

\pgfplotsset{compat=1.18}

\newcommand{\br}{\mathbf{r}}

\begin{document}

\title{Multipole expansion for dispersion forces -- watch this trace}
\author{Jivesh Kaushal}
\affiliation{Institut für Physik, Universität Rostock, Albert-Einstein-Straße 23-24, D-18059 Rostock, 
Germany}
\author{Chris Boldt}
\affiliation{Institut für Physik, Universität Rostock, Albert-Einstein-Straße 23-24, D-18059 Rostock, 
Germany}
\author{Stefan Scheel}
\email{stefan.scheel@uni-rostock.de}
\affiliation{Institut für Physik, Universität Rostock, Albert-Einstein-Straße 23-24, D-18059 Rostock, 
Germany}
\author{Athanasios Laliotis}
\affiliation{Laboratoire de Physique des Lasers, Universit{\'e} Sorbonne Paris Nord, F-93430, Villetaneuse, France}
\affiliation{CNRS, UMR 7538, LPL, 99 Avenue J.-B. Cl{\'e}ment, F-93430 Villetaneuse, France} 
\author{Paolo Pedri}
\email{paolo.pedri@univ-paris13.fr}
\affiliation{Laboratoire de Physique des Lasers, Universit{\'e} Sorbonne Paris Nord, F-93430, Villetaneuse, France}
\affiliation{CNRS, UMR 7538, LPL, 99 Avenue J.-B. Cl{\'e}ment, F-93430 Villetaneuse, France} 

\begin{abstract}
Light-matter interaction models invariably rely on the multipole expansion of the 
electromagnetic potentials generated by complex charge distributions. These multipoles are typically taken to be traceless, however, for a correct evaluation of dispersion forces at all distances, the validity of this assumption has to be checked carefully. Here, we revisit the concept of dispersion forces on an atom near a dielectric
surface from the perspective of macroscopic quantum electrodynamics
and find that, beyond the quadrupole, the multipoles cannot always be taken as fully
traceless. In particular, we show that the trace of the octupole moment contributes to 
Casimir-Polder interactions beyond the electrostatic regime.

\end{abstract}

\maketitle

\section{Introduction}

At the heart of any light-matter interaction lies the multipole expansion
\cite{amobook,llqmbook},  by which the Coulomb interaction potential at some 
observation point $\br$ is expanded in a Taylor series about the source charge at 
$\br'=\mathbf{0}$. In free space, this results in
\begin{multline}
V(\br - \br') = \frac{1}{\vert \br - \br' \vert} 
=\frac{1}{r} 
+ r_i'\left.\frac{\partial V}{\partial r_i'}\right\vert_{\br'=0} \\
+ \frac{r_i' r_j'}{2!}\left.\frac{\partial^2 V}{\partial r_i'\partial r_j'}\right
\vert_{\br'=0} 
+\frac{r_i' r_j' r_k'}{3!}\left.\frac{\partial^3 V}{\partial r_i'\partial 
r_j'\partial r_k'}\right\vert_{\br'=0} \\
+ \ldots, \label{eqn:V}
\end{multline}
with the derivatives at $\br'=\mathbf{0}$ encoding the structure of each multipole moment. 
As the Coulomb potential satisfies the Laplace equation, the multipoles can be chosen to 
be traceless. 
For example, for the quadrupole moment one finds that
\begin{equation}
    r_i'r_j' \left.\frac{\partial^2 V}{\partial r_i'\partial r_j'}
    \right\vert_{\br'=0} = 
    \left( r_i'r_j'-\frac{1}{3}\delta_{ij}r^{'2} \right) \left.
    \frac{\partial^2 V}{\partial r_i'\partial r_j'}
    \right\vert_{\br'=0}.
\end{equation}
This condition is equivalent to decomposing the angular momentum into direct sums 
of multipoles $m_{i_1,i_2,\ldots,i_\ell}$ of order $\ell$,
\begin{multline}
\frac{1}{\vert\boldsymbol{\br-\br'}\vert} = \sum_{\ell=0}^{\infty} 
m_{i_1,i_2,\ldots,i_\ell}\frac{r^{\prime \ell}}{r^{\ell+1}} \\
\times\sum_{m=-\ell}^{\ell}(-1)^m\sqrt{\frac{4\pi}{2\ell+1}}Y_{\ell m}(\theta,\phi);
\quad (r' < r).
\label{eqn:DS_Ylm}
\end{multline}

In the electrostatic regime, the traceless multipoles allow for a succinct, and in 
most situations, accurate calculation of a diverse array of physical interactions, 
ranging from the ubiquitous electric dipole-dipole interactions which are the 
foundation for characterizing physical and chemical properties of atoms and 
molecules \cite{amobook,craig1998,mQMbook,mQCbook}, to crystals in 
solid state physics \cite{sspbook,sspbook2}.
Once going outside the electrostatic regime,
one has to question the validity of the traceless 
condition for the multipole moments. A pertinent example regards dispersion 
interactions between atoms and surfaces, i.e. the Casimir--Polder interaction 
\cite{buhmann2012}
which, at least in the nonretarded limit, can be intuitively understood as the 
interaction of an atomic multipole with its surface-induced image, mediated
by ground-state fluctuations of the quantized electromagnetic field. 
Up to now, the main focus in the field of dispersion forces had been mostly on 
electric dipole moments. With the advent of experimental techniques to probe highly
excited Rydberg atoms near surfaces  \cite{sandoghdar1992, kuebler2010, rajasree2020, kaiser2022}, it has become necessary to include also higher-order multipoles in order to precisely describe atom-surface interactions \cite{crosse2010}. 

Recently, we have shown that higher-order multipole transitions will influence
measurements of the Casimir-Polder interaction of Rydberg atoms in 
microfabricated vapour cells \cite{dutta2024} using thin-cell transmission or reflection 
spectroscopy \cite{fichet2007,peyrot2019}
There, in vapour cells with submicrometer wall separation, the effects of quadrupole and octupole 
moments become larger than the expected measurement accuracy \cite{dutta2024}, thereby necessitating 
their inclusion in the modelling of the atom-surface interaction \cite{stourm2020}. 
{In Ref.~\cite{dutta2024}, it was shown that thin-cell spectroscopy can already register 
significant deviations from the electric dipole-electric dipole energy shifts for cell thicknesses starting at around $200~\text{nm}$, making it necessary to consider quadrupole-quadrupole effects, for which
an expression was derived in the non-retarded limit. This leaves open the question of multipole 
interactions just beyond the non-retarded regime. It was noted, that trace components from octupole 
interactions can also potentially contribute for distances comparable to (i.e.~just beyond) the 
non-retarded regime, even if it is exactly zero in the electrostatic limit.}

As early as the 1970s, there have been detailed theoretical analyses that 
question the implicit assumption of the traceless multipole moments 
\cite{raab1975,molqedbook,jenkins1994,salam1996,salam2000}, and the topic has remained active in one form or another to this day \cite{ostrovsky2006,salam2018,kosik2020}. In 
Ref.~\cite{raab1975}, the inconsistency in the description of magnetic (and 
electric) multipole moments for molecules is discussed in detail. The condition of 
tracelessness imposed on electric multipole moments beyond and including the 
quadrupole are found to be insufficient to describe the interaction energies 
associated with multipole coupling to an external field, and a non-traceless 
approach for all multipole orders is presented. This approach is further extended 
to the retarded (far-field) regime for chiral molecules \cite{jenkins1994}. 

In this article, we show the influence of multipole moments on dispersion forces.
We show that, outside the electrostatic regime, the assumption of fully traceless 
multipole moments is incorrect, and certain index contractions cannot be taken 
traceless. As a result, we discover a new term, emerging from octupole-dipole interactions. This contribution scales with the atom-surface distance $d$ as $1/d^6$ in the retarded limit, coinciding with the respective scaling for the quadraupole-quadrupole interaction. However, this new term only exists outside the non-retarded limit, which implies an emergence of this contribution at intermediate distances.

This article is organized as follows. 
In Sec.\ref{sec:MM_medium}, we show that, even for the Casimir-Polder interaction between 
an atom and a dielectric half-space, 
the use of traceless multipole moments is still permitted, at least in the electrostatic 
(nonretarded) limit. In Sec.~\ref{sec:retarded}, we then show an explicit example
for the appearance of trace contributions to the dipole-octupole Casimir--Polder 
interaction in the retarded limit.
Concluding remarks are given in Sec.~\ref{sec:conclusions}.

\section{Multipole dispersion {interaction} near a dielectric interface}
\label{sec:MM_medium}

In the physics of dispersion forces, one considers the interaction of an atom or
molecule with the medium-assisted quantized electromagnetic field, in which the
dielectric medium adds boundary conditions on the electromagnetic field at its
interface with free space. 
This is taken into account within the Green function expansion of the electromagnetic
fields that, besides the free-space Green function $\bm{G}^{(0)}$, also contains 
contributions $\bm{G}^{(1)}$ from the scattering off the matter-vacuum interface 
\cite{scheel2008,buhmann2012}.
The Casimir--Polder potential is then commonly derived using second-order perturbation 
theory using the interaction Hamiltonian
\begin{align}
    \hat{H}_\mathrm{int}&= -\hat{\mathbf{d}}\cdot\hat{\mathbf{E}}(\mathbf{r}_A)
    - \hat{\mathbf{Q}}\bullet\left[\bm{\nabla}\otimes
    \hat{\mathbf{E}}(\mathbf{r}_A)\right] \nonumber\\&- \hat{\bm{O}}\bullet\left[\bm{\nabla}\otimes\bm{\nabla}\hat{\mathbf{E}}(\mathbf{r_A})\right]\ldots
\end{align}
where $\bullet$ denotes the Frobenius inner product{, $\hat{\bm{d}}$, $\hat{\bm{Q}}$ and $\hat{\bm{O}}$ are the electric dipole, quadrupole and octupole operators respectively and $\hat{\bm{r}}_A$ the atomic position} \cite{crosse2010}. The interaction Hamiltonian represents an expansion into successively higher multipoles (dipoles, quadrupoles, octupoles etc.), contracted with derivatives of the electric field at the position of the atom.

Expanding the operator of the electric field in terms of dyadic Green functions 
\cite{dgfbookchento,scheel2008}, the Casimir--Polder potential associated with dipole-dipole interactions at zero temperature can be written as
\begin{equation}
    U_\mathrm{dip}(\mathbf{r}_A) = \frac{\hbar\mu_0}{2\pi} \int\limits_0^\infty
    d\xi\,\xi^2 \left[ \bm{\alpha}_{dd}^{(2)}(i\xi) \bullet 
    \bm{G}^{(1)}(\mathbf{r}_A,\mathbf{r}_A,i\xi) \right]
\end{equation}
with the dipole polarizability $\bm{\alpha}_{dd}^{(2)}(i\xi)$ and the scattering part 
$\bm{G}^{(1)}(\mathbf{r}_A,\mathbf{r}_A,i\xi)$ of the Green tensor, evaluated at
imaginary frequencies. 

Similarly, for quadrupole-quadrupole interaction, one finds \cite{crosse2010}
\begin{gather}
    U_\mathrm{quad}(\mathbf{r}_A) = \frac{\hbar\mu_0}{2\pi} \int\limits_0^\infty
    d\xi\,\xi^2 \left[ \bm{\alpha}_{qq}^{(4)}(i\xi) \bullet \right. \nonumber\\ \left.
    \bm{\nabla}\otimes\bm{G}^{(1)}(\mathbf{r}_A,\mathbf{r}_A,i\xi)\otimes
    \overleftarrow{\bm{\nabla}}\right]\,,\label{eq:Uqq}
\end{gather}
with the quadrupole polarizability
\begin{equation}
    \bm{\alpha}_{qq}^{(4)}(\omega) = \frac{1}{\hbar} \sum\limits_{k\ne n} \left[
    \frac{\mathbf{Q}_{nk}\otimes\mathbf{Q}_{kn}}{\omega_{kn}+\omega}
    +\frac{\mathbf{Q}_{nk}\otimes\mathbf{Q}_{kn}}{\omega_{kn}-\omega}\right]
\end{equation}
given in terms of the quadrupole transition moments 
$\mathbf{Q}_{nk}=\langle n|e(\hat{\mathbf{r}}\otimes\hat{\mathbf{r}})/2|k\rangle$ 
{and atomic transition frequencies $\omega_{kn}$}.
Note that the Frobenius product here acts as a generalized trace. Other types of 
multipole interactions can be read off by straightforward generalization. The 
number $l$ of gradient operators acting from the left/right on the Green tensor 
corresponds to the order $2^{l+1}$ of the multipole involved in the interaction.

Similar to free space, the multipoles in the Casimir--Polder interaction inherit
their traceless nature from the Green function, in case of dispersion interactions
specifically from the scattering part. Let us recall an important property of the
scattering Green tensor. The full Green tensor $\bm{G}=\bm{G}^{(0)}+\bm{G}^{(1)}$
obeys the inhomogeneous Helmholtz equation in the space including the dielectric
material, whereas the free-space Green tensor $\bm{G}^{(0)}$ obeys the 
inhomogeneous Helmholtz equation in free space. This means that the scattering part
$\bm{G}^{(1)}$ satisfies a homogeneous Helmholtz equation as it relates to a 
source-free problem, and is hence (double-sided) transverse, i.e. 
$\partial_iG^{(1)}_{ij}=G^{(1)}_{ij}\overleftarrow{\partial_j}=0$. 

This in turn means that any quadrupole-type interaction can be written 
using traceless quadrupole moments. To see this, let us decompose the quadrupole
tensor $Q_{ij}$ into a traceless part $\tilde{Q}_{ij}$ and its trace,
\begin{equation}
    Q_{ij} = \tilde{Q}_{ij} +\frac{1}{3}\mathrm{Tr}(\mathbf{Q}) \delta_{ij}\,.
\end{equation}
Hence, the contraction of the quadrupole moment with the gradient of the Green 
tensor, writing out the Frobenius product, is
\begin{equation}
    Q_{ij} \partial_i G^{(1)}_{jk} = \tilde{Q}_{ij} \partial_i G^{(1)}_{jk}
    +\frac{1}{3}\mathrm{Tr}(\mathbf{Q}) \partial_i G^{(1)}_{ik}
    = \tilde{Q}_{ij} \partial_i G^{(1)}_{jk}
\end{equation}
because the scattering Green tensor is divergence-free. Hence, all dispersion
interactions involving quadrupole moments can always be written using traceless quadrupole
tensors, usually defined by \cite{raab1975}
\begin{equation}
    \Theta_{ij} = \frac{1}{2} \left( 3 Q_{ij} -\mathrm{Tr}\mathbf{Q}\delta_{ij} 
    \right) = \frac{3}{2} \tilde{Q}_{ij} \,.
\end{equation}

Let us turn to the octupole moments, defined in their traceless form as 
\cite{raab1975}
\begin{equation}
    \Omega_{ijk} = \frac{1}{2} \left( 5O_{ijk} -O_{ill}\delta_{jk} -O_{jll}
    \delta_{ik} -O_{kll} \delta_{ij} \right)
\end{equation}
which are traceless in all pairs of tensor indices. Its contraction with suitably
chosen gradients of the scattering Green tensor then reads as
\begin{gather}
    \Omega_{ijk} \partial_i \partial_j G^{(1)}_{km} = 
    \frac{5}{2} O_{ijk} \partial_i \partial_j G^{(1)}_{km} \nonumber\\
    -\frac{1}{2} O_{ill} \partial_i \partial_j G^{(1)}_{jm} 
    -\frac{1}{2} O_{jll} \partial_i \partial_j G^{(1)}_{im} \nonumber \\
    -\frac{1}{2} O_{kll} \partial_i \partial_i G^{(1)}_{km} \,.
    \label{eq:tracelessoctupole}
\end{gather}
The second line in Eq.~(\ref{eq:tracelessoctupole}) again vanishes due to the 
transverse nature of the scattering Green tensor. However, the last term requires
special attention. Recalling that $\bm{G}^{(1)}$ satisfies the homogeneous 
Helmholtz equation, and is itself transverse, i.e. divergence-free, it satisfies
\begin{equation}
\label{eq:helmholtz}
    \Delta \bm{G}^{(1)}(\mathbf{r},\mathbf{r}',\omega) + \frac{\omega^2}{c^2}
    \varepsilon(\mathbf{r},\omega) \bm{G}^{(1)}(\mathbf{r},\mathbf{r}',\omega) 
    = \mathbf{0} \,.
\end{equation}
This means that the scattering part of the Green tensor does not fulfill the Laplace
equation as does the static Coulomb potential {\eqref{eqn:V}} above. Hence, the combination 
$\partial_i \partial_i G^{(1)}_{km}$ in Eq.~(\ref{eq:tracelessoctupole}) by itself 
does not immediately vanish, but it equates to $-k^2G^{(1)}_{km}$. 

Nevertheless, if we consider the 
electrostatic limit, where one defines the static Green tensor as
\begin{equation}
    \bm{\Gamma}(\mathbf{r},\mathbf{r}') = \lim\limits_{\omega\to0} 
    \frac{\omega^2}{c^2} \bm{G}^{(1)}(\mathbf{r},\mathbf{r}',\omega)\,,
\end{equation}
one indeed finds again that
\begin{equation}
    \Omega_{ijk} \partial_i \partial_j \Gamma_{km} = 
    \frac{5}{2} O_{ijk} \partial_i \partial_j \Gamma_{km} \ 
\end{equation}
This means that in the nonretarded, electrostatic limit, all octupole (and higher-order) moments can be chosen again to be traceless.

However, in the general case, i.e outside the limit of the electrostatic approximation, there is no immediate 
reason why the multipole moments beyond the quadrupole have to be traceless. This important conclusion
applies to ongoing experiments in which Casimir--Polder interactions are investigated 
at intermediate distances, and in which higher-order multipole interactions are 
important. This requires precision experiments, typically involving not too highly 
excited Rydberg atoms, at sub-millimeter distances from a surface. The Rydberg 
excitations ensure that the extent of the electronic wavefunction is no longer 
negligible, so that field gradients near the surface become important, and hence 
higher-order multipole interactions become relevant. This has to be balanced by the 
fact that the nonretarded regime extends further for larger principal quantum numbers 
$n$, as the dominant transitions have wavelengths that increase as $n^3$. Therefore, 
we can surmise that low-lying Rydberg atoms are a good choice to see these effects,
which points towards current experiments in nanocells \cite{dutta2024}. Additionally, higher order corrections could eventually 
also become of importance for metrological high precision measurements of Casimir-Polder 
or Casimir interactions looking for new forces \cite{laliotis2021review}.\\

\section{Trace contribution in the retarded limit}
\label{sec:retarded}

As shown above, the contraction of the (traceful) octupole moment tensor $O_{ijk}$ with 
the derivatives of the Green tensor gives
\begin{equation}
    O_{ijk} \partial_i \partial_j G^{(1)}_{km} 
    = \frac{2}{5} \Omega_{ijk} \partial_i \partial_j G^{(1)}_{km}
    -\frac{\omega^2}{5c^2} O_{kll} G^{(1)}_{km} 
\end{equation}
where, in the last term resulting from the trace of the octupole moment, we replaced the
wavenumber $k\mapsto\omega/c$. Note that this has the appearance of an effective dipole 
moment, which can be seen as follows. Take {for example} the $zzz$-component of the octupole tensor 
$e(\mathbf{r}\otimes\mathbf{r}\otimes\mathbf{r})/6$, its decomposition into 
spherical harmonics yields
\begin{gather}
    \frac{e}{6}(\mathbf{r}\otimes\mathbf{r}\otimes\mathbf{r})_{zzz} = 
    \frac{e}{6} r^3 \cos^3\Theta \nonumber \\
    = \frac{e}{6} r^3 \left[ \frac{4}{5} 
    \sqrt{\frac{\pi}{7}} Y_{3,0}(\Theta,\varphi) + \frac{6}{5} \sqrt{\frac{\pi}{3}}
    Y_{1,0}(\Theta,\varphi) \right]
\end{gather}
which contains, besides the expected octupole term (with $l=3$ in the spherical harmonic $Y_{lm}$), an effective dipole term (with $l=1$) that derives solely from the trace of the octupole moment. 
{Because of the spherical harmonic $Y_{1m}$, the selection rules are identical to those of a dipole, $\Delta l=\pm 1.$}

In second-order perturbation theory \cite{buhmann2012}, the trace contribution to the
energy shift is thus {(see Supplementary Material \cite{sm})}
\begin{gather}
    U_{od}^{\mathrm{trace}}(\mathbf{r}_A) = \nonumber \\
    \frac{\mu_0}{5\hbar c^2} \sum\limits_k \int\limits_0^\infty
    \frac{d\omega}{\omega_k+\omega} \omega^4 \mathbf{T}_{0k} \cdot 
    \mathrm{Im}\bm{G}^{(1)}(\mathbf{r}_A,\mathbf{r}_A,\omega)\cdot \mathbf{d}_{k0}
\end{gather}
where we have denoted $(\mathbf{T}_{0k})_i\equiv \langle 0\vert \hat{O}_{ill} \vert k\rangle$ for the transition $0\to k$.
This results in a contribution to the Casimir--Polder shift as
\begin{gather}
    U_{od}^{\mathrm{trace}}(\mathbf{r}_A) = \nonumber \\
    \frac{\hbar\mu_0}{10\pi c^2} \int\limits_0^\infty d\xi\,\xi^4 \mathrm{Tr}\left[
    \bm{\alpha}_{od}^{(4)}(i\xi)\cdot\bm{G}^{(1)}(\mathbf{r}_A,\mathbf{r}_A,i\xi)\right]\,.
\end{gather}
Note that the replacement $\omega\mapsto i\xi$ leads, via the extra two powers
of $\omega$, to a sign change. We have further defined an octupole-dipole polarisability
\begin{equation}
\bm{\alpha}_{od}^{(4)}(i\xi) = \frac{1}{\hbar} \sum\limits_k \left( 
\frac{\mathbf{d}_{k0}\otimes\mathbf{T}_{0k}}{i\xi+\omega_k} -
\frac{\mathbf{T}_{0k}\otimes\mathbf{d}_{k0}}{i\xi-\omega_k} \right)
\end{equation}
with the effective dipole moment resulting from the octupole trace.
{Similarly, a dipole-octupole interaction can be derived in the same way with an analogous dipole-octupole polarizability (see Supplementary Material \cite{sm}).}


Let us now {consider an atom near a planar interface and}  recall how the scattering Green 
function for a half-space {with the interface at $z=0$} can be 
conveniently written, and how the retarded limit of the Casimir--Polder interaction can be 
subsequently derived.

Using the Fresnel reflection coefficients, the scattering Green tensor in the coincidence limit is given by \cite{scheel2008,buhmann2012}
\begin{gather}
    \bm{G}^{(1)}(\mathbf{r},\mathbf{r},\omega) = \frac{i}{8\pi} \int\limits_0^\infty dq 
    \frac{q}{\beta_0} e^{2i\beta_0 z_A} \nonumber \\ \times
    \left[ r_s \begin{pmatrix} 1&0&0\\0&1&0\\0&0&0 \end{pmatrix} + r_p \frac{c^2}{\omega^2}
    \begin{pmatrix} -\beta_0^2&0&0\\0&-\beta_0^2&0\\0&0&2q^2 \end{pmatrix}
    \right]
\end{gather}
where $\beta_0=\sqrt{\omega^2/c^2-q^2}$ $[\mathrm{Im}\beta_0\ge 0]$ is the $z$-component 
of the wavenumber in free space, and $q$ the component perpendicular to it. 
Rotating to 
the imaginary frequency axis and defining $b_0=\sqrt{\xi^2/c^2+q^2}$, one finds the 
scattering Green tensor at imaginary frequencies as \cite{buhmann2012}
\begin{gather}
    \bm{G}^{(1)}(\mathbf{r},\mathbf{r},i\xi) = \frac{1}{8\pi} \int\limits_{\xi/c}^\infty db_0 
    \,e^{-2b_0 z_A} \nonumber \\ \times
    \left[ r_s \begin{pmatrix} 1&0&0\\0&1&0\\0&0&0 \end{pmatrix} - r_p \frac{c^2}{\xi^2}
    \begin{pmatrix} b_0^2&0&0\\0&b_0^2&0\\0&0&2b_0^2-2\frac{\xi^2}{c^2} \end{pmatrix}
    \right] \,.
\end{gather}

In the retarded limit, one assumes that the atom-surface distance $z_A$ dominates over 
all transition wavelengths. 
One can therefore approximate the atomic polarisability as well as the optical properties of the 
half-space by their static values, i.e. $\bm{\alpha}_{od}^{(4)}(i\xi)\simeq\bm{\alpha}_{od}(0)$ and 
$\varepsilon(i\xi)\simeq\varepsilon(0)\equiv\varepsilon$ \cite{scheel2008,buhmann2012}.

Changing variables to $v=b_0c/\xi$, the order of integration in the Casimir--Polder 
potential can be changed to
\begin{equation}
    \int\limits_0^\infty d\xi \int\limits_{\xi/c}^\infty db_0 \mapsto \int\limits_1^\infty dv
    \int\limits_0^\infty d\xi \,\frac{\xi}{c}\,,
\end{equation}
and integrating over $\xi$ yields
\begin{equation}
    \int\limits_0^\infty d\xi \,\xi^5 \, e^{-2\xi vz_A/c} = \frac{15}{8} \left( \frac{c}{vz_A}
    \right)^6 \,.
\end{equation}
Combining everything yields the contribution of the octupole trace to the octupole-dipole
Casimir--Polder energy in the retarded limit as
\begin{gather}
    U_{od}^{\mathrm{trace}}(z_A) \simeq \frac{3\hbar c}{128\pi^2\varepsilon_0z_A^6} 
    \int\limits_1^\infty \frac{dv}{v^6} 
    \nonumber \\ \times \mathrm{Tr} \left[ \bm{\alpha}_{od}(0) \cdot
    \left\{ r_s \begin{pmatrix} 1&0&0\\0&1&0\\0&0&0 \end{pmatrix} -r_p
    \begin{pmatrix} v^2&0&0\\0&v^2&0\\0&0&2v^2-2 \end{pmatrix} \right\} \right]
\end{gather}
where the Fresnel reflection coefficients are approximated by
\begin{equation}
    r_s\simeq\frac{v-\sqrt{\varepsilon-1+v^2}}{v+\sqrt{\varepsilon-1+v^2}} \,,\quad
    r_p\simeq\frac{\varepsilon v-\sqrt{\varepsilon-1+v^2}}{\varepsilon v+\sqrt{\varepsilon-1+v^2}} 
    \,.
\end{equation}
In the limit of a perfectly reflecting half-space with $r_s=-1$ and $r_p=1$, one can even 
perform the $v$-integral to obtain
\begin{equation}
\label{eq:Uod}
    U_{od}^{\mathrm{trace}}(z_A) \simeq -\frac{\hbar c}{160\pi^2\varepsilon_0 z_A^6}
    \mathrm{Tr} \left[ \bm{\alpha}_{od}(0) \cdot \begin{pmatrix} 2&0&0\\0&2&0\\0&0&1 \end{pmatrix} 
    \right] \,.
\end{equation}

As expected, on dimensional grounds, the potential falls off as $z_A^{-6}$, despite the 
fact that the trace part of the octupole moment behaves as an effective dipole. 
This is the main result of this article that demonstrates a nonzero dipole-octupole
contribution to the Casimir-Polder interaction in the retarded regime.

{We now compare the trace contribution to the quadrupole-quadrupole interaction in order
to estimate its potential impact. For that, we return to Eq.~(\ref{eq:Uqq}) and compute the 
quadrupole-quadrupole interaction in the retarded limit, and perform a rotational averaging.
Details of this calculation are given in the Supplementary Material, Sec.~II \cite{sm}. The result is
\begin{equation}
    U_{qq}(z_A) \simeq -\frac{5\hbar c}{64\pi^2\varepsilon_0 z_A^6} \alpha_{qq}(0)    
\end{equation}
with the static, isotropic quadrupole polarizability $\alpha_{qq}(0)$. Performing a similar isotropic
averaging to Eq.~(\ref{eq:Uqq}), we find for the ratio between the dipole-octupole trace and
the quadrupole-quadrupole interaction in the retarded limit the simple result

\begin{equation}
    \frac{U_{od} + U_{do}}{U_{qq}}=\frac{2}{15} \frac{\alpha_{od}(0) + \alpha_{do}(0)}{\alpha_{qq}(0)}
\end{equation}

showing that the trace part of the dipole-octupole interaction could be on the same order
of magnitude as the quadrupole-quadrupole interaction.
Indeed, taking the example of the $|6S_{1/2}\rangle$ ground state of cesium, the 
dipole-octupole energy shift amounts to approximately half the quadrupole-quadrupole shift {(further details in the Supplementary Material \cite{sm})}.

{We also note here, that while the magnetic dipole and the electric quadrupole interaction arise at the same level of perturbation theory, for atomic multipole interactions one has to also take into account how the relevant multipole transition moments scale with atomic properties. Electric multipole transition moments scale strongly with the principal quantum $n$, for example dipole moments $d\propto n^2$ or quadrupole moments $q\propto n^4$ and so on. In  contrast, magnetic transition moments are (almost) independent of the principal quantum number, as the magnetic moment operator $\hat{\mathbf{L}}+2\hat{\mathbf{S}}$ does not affect the radial part of the atomic wavefunction. Thus, for the Rydberg atoms envisaged in our manuscript, electric multipole transitions will dominate over possible magnetic dipole transitions.}
}

\section{Conclusions}
\label{sec:conclusions}

We have reconsidered the pertinent question of using traceless multipole moments
in the computation of higher-order multipole dispersion forces. We have shown that
the atomic multipole moments can still be chosen to be traceless in the nonretarded (or 
electrostatic) limit. This is due to the fact that the scattering part of the Green 
tensor, that is responsible for the dispersion interaction, is transverse and, in the 
electrostatic limit, also obeys the Laplace equation which is crucial for the appearance 
of the traceless moments.

However, we also demonstrate that outside the nonretarded limit, where the scattering part of the Green tensor merely solves the homogeneous Helmholtz equation, not all index pairs of
the octupole tensor give a traceless contribution. We furthermore explicitly calculate this trace contribution in the retarded limit.  
Finally, we have outlined some possible experimental scenarios such as precision spectroscopy 
of Rydberg atoms confined in thin vapor cells or high-precision fundamental physics 
measurements and metrology, where the higher-order corrections discussed here could become 
relevant.

\bibliographystyle{apsrev4-1}
\bibliography{Main.bib}

\begin{thebibliography}{29}%
\makeatletter
\providecommand \@ifxundefined [1]{%
 \@ifx{#1\undefined}
}%
\providecommand \@ifnum [1]{%
 \ifnum #1\expandafter \@firstoftwo
 \else \expandafter \@secondoftwo
 \fi
}%
\providecommand \@ifx [1]{%
 \ifx #1\expandafter \@firstoftwo
 \else \expandafter \@secondoftwo
 \fi
}%
\providecommand \natexlab [1]{#1}%
\providecommand \enquote  [1]{``#1''}%
\providecommand \bibnamefont  [1]{#1}%
\providecommand \bibfnamefont [1]{#1}%
\providecommand \citenamefont [1]{#1}%
\providecommand \href@noop [0]{\@secondoftwo}%
\providecommand \href [0]{\begingroup \@sanitize@url \@href}%
\providecommand \@href[1]{\@@startlink{#1}\@@href}%
\providecommand \@@href[1]{\endgroup#1\@@endlink}%
\providecommand \@sanitize@url [0]{\catcode `\\12\catcode `\$12\catcode
  `\&12\catcode `\#12\catcode `\^12\catcode `\_12\catcode `\%12\relax}%
\providecommand \@@startlink[1]{}%
\providecommand \@@endlink[0]{}%
\providecommand \url  [0]{\begingroup\@sanitize@url \@url }%
\providecommand \@url [1]{\endgroup\@href {#1}{\urlprefix }}%
\providecommand \urlprefix  [0]{URL }%
\providecommand \Eprint [0]{\href }%
\providecommand \doibase [0]{http://dx.doi.org/}%
\providecommand \selectlanguage [0]{\@gobble}%
\providecommand \bibinfo  [0]{\@secondoftwo}%
\providecommand \bibfield  [0]{\@secondoftwo}%
\providecommand \translation [1]{[#1]}%
\providecommand \BibitemOpen [0]{}%
\providecommand \bibitemStop [0]{}%
\providecommand \bibitemNoStop [0]{.\EOS\space}%
\providecommand \EOS [0]{\spacefactor3000\relax}%
\providecommand \BibitemShut  [1]{\csname bibitem#1\endcsname}%
\let\auto@bib@innerbib\@empty
\bibitem [{\citenamefont {Brandsen}\ and\ \citenamefont
  {Joachen}(2003)}]{amobook}%
  \BibitemOpen
  \bibfield  {author} {\bibinfo {author} {\bibfnamefont {B.~H.}\ \bibnamefont
  {Brandsen}}\ and\ \bibinfo {author} {\bibfnamefont {C.~J.}\ \bibnamefont
  {Joachen}},\ }\href@noop {} {\emph {\bibinfo {title} {Physics of Atoms and
  Molecules}}},\ \bibinfo {edition} {2nd}\ ed.\ (\bibinfo  {publisher} {Pearson
  Education},\ \bibinfo {year} {2003})\BibitemShut {NoStop}%
\bibitem [{\citenamefont {Landau}\ and\ \citenamefont
  {Lifshitz}(1981)}]{llqmbook}%
  \BibitemOpen
  \bibfield  {author} {\bibinfo {author} {\bibfnamefont {L.~D.}\ \bibnamefont
  {Landau}}\ and\ \bibinfo {author} {\bibfnamefont {E.~M.}\ \bibnamefont
  {Lifshitz}},\ }\href@noop {} {\emph {\bibinfo {title} {Quantum Mechanics:
  Non-Relativistic}}},\ \bibinfo {edition} {3rd}\ ed.\ (\bibinfo  {publisher}
  {Butterworth-Heineman},\ \bibinfo {year} {1981})\BibitemShut {NoStop}%
\bibitem [{\citenamefont {Craig}\ and\ \citenamefont
  {Thirunamachandran}(1998)}]{craig1998}%
  \BibitemOpen
  \bibfield  {author} {\bibinfo {author} {\bibfnamefont {D.~P.}\ \bibnamefont
  {Craig}}\ and\ \bibinfo {author} {\bibfnamefont {T.}~\bibnamefont
  {Thirunamachandran}},\ }\href@noop {} {\emph {\bibinfo {title} {Molecular
  Quantum Electrodynamics}}}\ (\bibinfo  {publisher} {Dover Publications},\
  \bibinfo {year} {1998})\BibitemShut {NoStop}%
\bibitem [{\citenamefont {Atkins}\ and\ \citenamefont
  {Friedman}(2010)}]{mQMbook}%
  \BibitemOpen
  \bibfield  {author} {\bibinfo {author} {\bibfnamefont {P.~W.}\ \bibnamefont
  {Atkins}}\ and\ \bibinfo {author} {\bibfnamefont {R.~S.}\ \bibnamefont
  {Friedman}},\ }\href@noop {} {\emph {\bibinfo {title} {Molecular Quantum
  Mechanics}}}\ (\bibinfo  {publisher} {Oxford University Press},\ \bibinfo
  {year} {2010})\BibitemShut {NoStop}%
\bibitem [{\citenamefont {Haken}\ \emph {et~al.}(2004)\citenamefont {Haken},
  \citenamefont {Wolf},\ and\ \citenamefont {Brewer}}]{mQCbook}%
  \BibitemOpen
  \bibfield  {author} {\bibinfo {author} {\bibfnamefont {H.}~\bibnamefont
  {Haken}}, \bibinfo {author} {\bibfnamefont {H.~C.}\ \bibnamefont {Wolf}}, \
  and\ \bibinfo {author} {\bibfnamefont {W.~D.}\ \bibnamefont {Brewer}},\
  }\href@noop {} {\emph {\bibinfo {title} {Molecular Physics and Elements of
  Quantum Chemistry}}}\ (\bibinfo  {publisher} {Springer},\ \bibinfo {year}
  {2004})\BibitemShut {NoStop}%
\bibitem [{\citenamefont {Peierls}(1955)}]{sspbook}%
  \BibitemOpen
  \bibfield  {author} {\bibinfo {author} {\bibfnamefont {R.~E.}\ \bibnamefont
  {Peierls}},\ }\href@noop {} {\emph {\bibinfo {title} {Quantum Theory of
  Solids}}}\ (\bibinfo  {publisher} {Clarendon Press},\ \bibinfo {year}
  {1955})\BibitemShut {NoStop}%
\bibitem [{\citenamefont {Kittel}(1991)}]{sspbook2}%
  \BibitemOpen
  \bibfield  {author} {\bibinfo {author} {\bibfnamefont {C.~S.}\ \bibnamefont
  {Kittel}},\ }\href@noop {} {\emph {\bibinfo {title} {Introduction to Solid
  State Physics}}}\ (\bibinfo  {publisher} {John Wiley \& Sons},\ \bibinfo
  {year} {1991})\BibitemShut {NoStop}%
\bibitem [{\citenamefont {Buhmann}(2012)}]{buhmann2012}%
  \BibitemOpen
  \bibfield  {author} {\bibinfo {author} {\bibfnamefont {S.~Y.}\ \bibnamefont
  {Buhmann}},\ }\href@noop {} {\emph {\bibinfo {title} {Dispersion Forces Vol.
  I and II}}}\ (\bibinfo  {publisher} {Springer Publications},\ \bibinfo {year}
  {2012})\BibitemShut {NoStop}%
\bibitem [{\citenamefont {Sandoghdar}\ \emph {et~al.}(1992)\citenamefont
  {Sandoghdar}, \citenamefont {Sukenik}, \citenamefont {Hinds},\ and\
  \citenamefont {Haroche}}]{sandoghdar1992}%
  \BibitemOpen
  \bibfield  {author} {\bibinfo {author} {\bibfnamefont {V.}~\bibnamefont
  {Sandoghdar}}, \bibinfo {author} {\bibfnamefont {C.~I.}\ \bibnamefont
  {Sukenik}}, \bibinfo {author} {\bibfnamefont {E.~A.}\ \bibnamefont {Hinds}},
  \ and\ \bibinfo {author} {\bibfnamefont {S.}~\bibnamefont {Haroche}},\
  }\href@noop {} {\bibfield  {journal} {\bibinfo  {journal} {Phys. Rev. Lett.}\
  }\textbf {\bibinfo {volume} {68}},\ \bibinfo {pages} {3432} (\bibinfo {year}
  {1992})}\BibitemShut {NoStop}%
\bibitem [{\citenamefont {K{\"u}bler}\ \emph {et~al.}(2010)\citenamefont
  {K{\"u}bler}, \citenamefont {Shaffer}, \citenamefont {Baluktsian},
  \citenamefont {L{\"o}w},\ and\ \citenamefont {Pfau}}]{kuebler2010}%
  \BibitemOpen
  \bibfield  {author} {\bibinfo {author} {\bibfnamefont {H.}~\bibnamefont
  {K{\"u}bler}}, \bibinfo {author} {\bibfnamefont {J.~P.}\ \bibnamefont
  {Shaffer}}, \bibinfo {author} {\bibfnamefont {T.}~\bibnamefont {Baluktsian}},
  \bibinfo {author} {\bibfnamefont {R.}~\bibnamefont {L{\"o}w}}, \ and\
  \bibinfo {author} {\bibfnamefont {T.}~\bibnamefont {Pfau}},\ }\href@noop {}
  {\bibfield  {journal} {\bibinfo  {journal} {Nature Photonics}\ }\textbf
  {\bibinfo {volume} {4}},\ \bibinfo {pages} {112} (\bibinfo {year}
  {2010})}\BibitemShut {NoStop}%
\bibitem [{\citenamefont {Rajasree}\ \emph {et~al.}(2020)\citenamefont
  {Rajasree}, \citenamefont {Ray}, \citenamefont {Karlsson}, \citenamefont
  {Everett},\ and\ \citenamefont {Chormaic}}]{rajasree2020}%
  \BibitemOpen
  \bibfield  {author} {\bibinfo {author} {\bibfnamefont {K.~S.}\ \bibnamefont
  {Rajasree}}, \bibinfo {author} {\bibfnamefont {T.}~\bibnamefont {Ray}},
  \bibinfo {author} {\bibfnamefont {K.}~\bibnamefont {Karlsson}}, \bibinfo
  {author} {\bibfnamefont {J.~L.}\ \bibnamefont {Everett}}, \ and\ \bibinfo
  {author} {\bibfnamefont {S.~N.}\ \bibnamefont {Chormaic}},\ }\href@noop {}
  {\bibfield  {journal} {\bibinfo  {journal} {Phys. Rev. Res.}\ }\textbf
  {\bibinfo {volume} {2}},\ \bibinfo {pages} {012038} (\bibinfo {year}
  {2020})}\BibitemShut {NoStop}%
\bibitem [{\citenamefont {Kaiser}\ \emph {et~al.}(2022)\citenamefont {Kaiser},
  \citenamefont {Glaser}, \citenamefont {Ley}, \citenamefont {Grimmel},
  \citenamefont {Hattermann}, \citenamefont {Bothner}, \citenamefont {Koelle},
  \citenamefont {Kleiner}, \citenamefont {Petrosyan}, \citenamefont
  {G{\"u}nther},\ and\ \citenamefont {Fort{\'a}gh}}]{kaiser2022}%
  \BibitemOpen
  \bibfield  {author} {\bibinfo {author} {\bibfnamefont {M.}~\bibnamefont
  {Kaiser}}, \bibinfo {author} {\bibfnamefont {C.}~\bibnamefont {Glaser}},
  \bibinfo {author} {\bibfnamefont {L.~Y.}\ \bibnamefont {Ley}}, \bibinfo
  {author} {\bibfnamefont {J.}~\bibnamefont {Grimmel}}, \bibinfo {author}
  {\bibfnamefont {H.}~\bibnamefont {Hattermann}}, \bibinfo {author}
  {\bibfnamefont {D.}~\bibnamefont {Bothner}}, \bibinfo {author} {\bibfnamefont
  {D.}~\bibnamefont {Koelle}}, \bibinfo {author} {\bibfnamefont
  {R.}~\bibnamefont {Kleiner}}, \bibinfo {author} {\bibfnamefont
  {D.}~\bibnamefont {Petrosyan}}, \bibinfo {author} {\bibfnamefont
  {A.}~\bibnamefont {G{\"u}nther}}, \ and\ \bibinfo {author} {\bibfnamefont
  {J.}~\bibnamefont {Fort{\'a}gh}},\ }\href@noop {} {\bibfield  {journal}
  {\bibinfo  {journal} {Phys. Rev. Research}\ }\textbf {\bibinfo {volume}
  {4}},\ \bibinfo {pages} {013207} (\bibinfo {year} {2022})}\BibitemShut
  {NoStop}%
\bibitem [{\citenamefont {Crosse}\ \emph {et~al.}(2010)\citenamefont {Crosse},
  \citenamefont {\r{A}. Ellingsen}, \citenamefont {Clements}, \citenamefont
  {Buhmann},\ and\ \citenamefont {Scheel}}]{crosse2010}%
  \BibitemOpen
  \bibfield  {author} {\bibinfo {author} {\bibfnamefont {J.~A.}\ \bibnamefont
  {Crosse}}, \bibinfo {author} {\bibfnamefont {S.}~\bibnamefont {\r{A}.
  Ellingsen}}, \bibinfo {author} {\bibfnamefont {K.}~\bibnamefont {Clements}},
  \bibinfo {author} {\bibfnamefont {S.~Y.}\ \bibnamefont {Buhmann}}, \ and\
  \bibinfo {author} {\bibfnamefont {S.}~\bibnamefont {Scheel}},\ }\href@noop {}
  {\bibfield  {journal} {\bibinfo  {journal} {Phys. Rev. A}\ }\textbf {\bibinfo
  {volume} {82}},\ \bibinfo {pages} {019101(R)} (\bibinfo {year}
  {2010})}\BibitemShut {NoStop}%
\bibitem [{\citenamefont {Dutta}\ \emph {et~al.}(2024)\citenamefont {Dutta},
  \citenamefont {de~Aquino~Carvalho}, \citenamefont {Garcia-Arellano},
  \citenamefont {Pedri}, \citenamefont {Laliotis}, \citenamefont {Boldt},
  \citenamefont {Kaushal},\ and\ \citenamefont {Scheel}}]{dutta2024}%
  \BibitemOpen
  \bibfield  {author} {\bibinfo {author} {\bibfnamefont {B.}~\bibnamefont
  {Dutta}}, \bibinfo {author} {\bibfnamefont {J.~C.}\ \bibnamefont
  {de~Aquino~Carvalho}}, \bibinfo {author} {\bibfnamefont {G.}~\bibnamefont
  {Garcia-Arellano}}, \bibinfo {author} {\bibfnamefont {P.}~\bibnamefont
  {Pedri}}, \bibinfo {author} {\bibfnamefont {A.}~\bibnamefont {Laliotis}},
  \bibinfo {author} {\bibfnamefont {C.}~\bibnamefont {Boldt}}, \bibinfo
  {author} {\bibfnamefont {J.}~\bibnamefont {Kaushal}}, \ and\ \bibinfo
  {author} {\bibfnamefont {S.}~\bibnamefont {Scheel}},\ }\href@noop {}
  {\bibfield  {journal} {\bibinfo  {journal} {Phys. Rev. Research}\ }\textbf
  {\bibinfo {volume} {6}},\ \bibinfo {pages} {L022035} (\bibinfo {year}
  {2024})}\BibitemShut {NoStop}%
\bibitem [{\citenamefont {Fichet}\ \emph {et~al.}(2007)\citenamefont {Fichet}
  \emph {et~al.}}]{fichet2007}%
  \BibitemOpen
  \bibfield  {author} {\bibinfo {author} {\bibfnamefont {M.}~\bibnamefont
  {Fichet}} \emph {et~al.},\ }\href@noop {} {\bibfield  {journal} {\bibinfo
  {journal} {Europhys. Lett.}\ }\textbf {\bibinfo {volume} {77}},\ \bibinfo
  {pages} {54001} (\bibinfo {year} {2007})}\BibitemShut {NoStop}%
\bibitem [{\citenamefont {Peyrot}\ \emph {et~al.}(2019)\citenamefont {Peyrot},
  \citenamefont {{\v{S}}ibali{\'c}}, \citenamefont {Sortais}, \citenamefont
  {Browaeys}, \citenamefont {Sargsyan}, \citenamefont {Sarkisyan},
  \citenamefont {Hughes},\ and\ \citenamefont {Adams}}]{peyrot2019}%
  \BibitemOpen
  \bibfield  {author} {\bibinfo {author} {\bibfnamefont {T.}~\bibnamefont
  {Peyrot}}, \bibinfo {author} {\bibfnamefont {N.}~\bibnamefont
  {{\v{S}}ibali{\'c}}}, \bibinfo {author} {\bibfnamefont {Y.~R.~P.}\
  \bibnamefont {Sortais}}, \bibinfo {author} {\bibfnamefont {A.}~\bibnamefont
  {Browaeys}}, \bibinfo {author} {\bibfnamefont {A.}~\bibnamefont {Sargsyan}},
  \bibinfo {author} {\bibfnamefont {D.}~\bibnamefont {Sarkisyan}}, \bibinfo
  {author} {\bibfnamefont {I.~G.}\ \bibnamefont {Hughes}}, \ and\ \bibinfo
  {author} {\bibfnamefont {C.~S.}\ \bibnamefont {Adams}},\ }\href@noop {}
  {\bibfield  {journal} {\bibinfo  {journal} {Phys. Rev. A}\ }\textbf {\bibinfo
  {volume} {100}},\ \bibinfo {pages} {022503} (\bibinfo {year}
  {2019})}\BibitemShut {NoStop}%
\bibitem [{\citenamefont {Stourm}\ \emph {et~al.}(2020)\citenamefont {Stourm},
  \citenamefont {Lepers}, \citenamefont {Robert}, \citenamefont {Nic~Chormaic},
  \citenamefont {M{\o}lmer},\ and\ \citenamefont {Brion}}]{stourm2020}%
  \BibitemOpen
  \bibfield  {author} {\bibinfo {author} {\bibfnamefont {E.}~\bibnamefont
  {Stourm}}, \bibinfo {author} {\bibfnamefont {M.}~\bibnamefont {Lepers}},
  \bibinfo {author} {\bibfnamefont {J.}~\bibnamefont {Robert}}, \bibinfo
  {author} {\bibfnamefont {S.}~\bibnamefont {Nic~Chormaic}}, \bibinfo {author}
  {\bibfnamefont {K.}~\bibnamefont {M{\o}lmer}}, \ and\ \bibinfo {author}
  {\bibfnamefont {E.}~\bibnamefont {Brion}},\ }\href@noop {} {\bibfield
  {journal} {\bibinfo  {journal} {Phys. Rev. A}\ }\textbf {\bibinfo {volume}
  {101}},\ \bibinfo {pages} {052508} (\bibinfo {year} {2020})}\BibitemShut
  {NoStop}%
\bibitem [{\citenamefont {Raab}(1975)}]{raab1975}%
  \BibitemOpen
  \bibfield  {author} {\bibinfo {author} {\bibfnamefont {R.~E.}\ \bibnamefont
  {Raab}},\ }\href@noop {} {\bibfield  {journal} {\bibinfo  {journal} {Mol.
  Phys.}\ }\textbf {\bibinfo {volume} {29}},\ \bibinfo {pages} {1323} (\bibinfo
  {year} {1975})}\BibitemShut {NoStop}%
\bibitem [{\citenamefont {Craig}\ and\ \citenamefont
  {Thirunamachandran}(1984)}]{molqedbook}%
  \BibitemOpen
  \bibfield  {author} {\bibinfo {author} {\bibfnamefont {D.~P.}\ \bibnamefont
  {Craig}}\ and\ \bibinfo {author} {\bibfnamefont {T.}~\bibnamefont
  {Thirunamachandran}},\ }\href@noop {} {\emph {\bibinfo {title} {Molecular
  Quantum Electrodynamics}}}\ (\bibinfo  {publisher} {Academic Press Inc},\
  \bibinfo {year} {1984})\BibitemShut {NoStop}%
\bibitem [{\citenamefont {Jenkins}\ \emph {et~al.}(1994)\citenamefont
  {Jenkins}, \citenamefont {Salam},\ and\ \citenamefont
  {Thirunamachandran}}]{jenkins1994}%
  \BibitemOpen
  \bibfield  {author} {\bibinfo {author} {\bibfnamefont {J.~K.}\ \bibnamefont
  {Jenkins}}, \bibinfo {author} {\bibfnamefont {A.}~\bibnamefont {Salam}}, \
  and\ \bibinfo {author} {\bibfnamefont {T.}~\bibnamefont
  {Thirunamachandran}},\ }\href@noop {} {\bibfield  {journal} {\bibinfo
  {journal} {Phys. Rev. A}\ }\textbf {\bibinfo {volume} {50}},\ \bibinfo
  {pages} {4767} (\bibinfo {year} {1994})}\BibitemShut {NoStop}%
\bibitem [{\citenamefont {Salam}\ and\ \citenamefont
  {Thirunamachandran}(1996)}]{salam1996}%
  \BibitemOpen
  \bibfield  {author} {\bibinfo {author} {\bibfnamefont {A.}~\bibnamefont
  {Salam}}\ and\ \bibinfo {author} {\bibfnamefont {T.}~\bibnamefont
  {Thirunamachandran}},\ }\href@noop {} {\bibfield  {journal} {\bibinfo
  {journal} {J. Phys. B}\ }\textbf {\bibinfo {volume} {104}},\ \bibinfo {pages}
  {5094} (\bibinfo {year} {1996})}\BibitemShut {NoStop}%
\bibitem [{\citenamefont {Salam}(2000)}]{salam2000}%
  \BibitemOpen
  \bibfield  {author} {\bibinfo {author} {\bibfnamefont {A.}~\bibnamefont
  {Salam}},\ }\href@noop {} {\bibfield  {journal} {\bibinfo  {journal} {J.
  Phys. B}\ }\textbf {\bibinfo {volume} {33}},\ \bibinfo {pages} {2181}
  (\bibinfo {year} {2000})}\BibitemShut {NoStop}%
\bibitem [{\citenamefont {Ostrovsky}\ \emph {et~al.}(2006)\citenamefont
  {Ostrovsky}, \citenamefont {Vrinceau},\ and\ \citenamefont
  {Flannery}}]{ostrovsky2006}%
  \BibitemOpen
  \bibfield  {author} {\bibinfo {author} {\bibfnamefont {V.~N.}\ \bibnamefont
  {Ostrovsky}}, \bibinfo {author} {\bibfnamefont {D.}~\bibnamefont {Vrinceau}},
  \ and\ \bibinfo {author} {\bibfnamefont {M.~R.}\ \bibnamefont {Flannery}},\
  }\href@noop {} {\bibfield  {journal} {\bibinfo  {journal} {Phys. Rev. A}\
  }\textbf {\bibinfo {volume} {74}},\ \bibinfo {pages} {022720} (\bibinfo
  {year} {2006})}\BibitemShut {NoStop}%
\bibitem [{\citenamefont {Salam}(2018)}]{salam2018}%
  \BibitemOpen
  \bibfield  {author} {\bibinfo {author} {\bibfnamefont {A.}~\bibnamefont
  {Salam}},\ }\href@noop {} {\bibfield  {journal} {\bibinfo  {journal} {Mol.
  Phys.}\ }\textbf {\bibinfo {volume} {117}},\ \bibinfo {pages} {2217}
  (\bibinfo {year} {2018})}\BibitemShut {NoStop}%
\bibitem [{\citenamefont {Kosik}\ \emph {et~al.}(2020)\citenamefont {Kosik},
  \citenamefont {Burlayenko}, \citenamefont {Rockstuhl}, \citenamefont
  {Fernandez-Corbaton},\ and\ \citenamefont {Slowik}}]{kosik2020}%
  \BibitemOpen
  \bibfield  {author} {\bibinfo {author} {\bibfnamefont {M.}~\bibnamefont
  {Kosik}}, \bibinfo {author} {\bibfnamefont {O.}~\bibnamefont {Burlayenko}},
  \bibinfo {author} {\bibfnamefont {C.}~\bibnamefont {Rockstuhl}}, \bibinfo
  {author} {\bibfnamefont {I.}~\bibnamefont {Fernandez-Corbaton}}, \ and\
  \bibinfo {author} {\bibfnamefont {K.}~\bibnamefont {Slowik}},\ }\href@noop {}
  {\bibfield  {journal} {\bibinfo  {journal} {Sci. Reports}\ }\textbf {\bibinfo
  {volume} {10}},\ \bibinfo {pages} {5879} (\bibinfo {year}
  {2020})}\BibitemShut {NoStop}%
\bibitem [{\citenamefont {Scheel}\ and\ \citenamefont
  {Buhmann}(2008)}]{scheel2008}%
  \BibitemOpen
  \bibfield  {author} {\bibinfo {author} {\bibfnamefont {S.}~\bibnamefont
  {Scheel}}\ and\ \bibinfo {author} {\bibfnamefont {S.~Y.}\ \bibnamefont
  {Buhmann}},\ }\href@noop {} {\bibfield  {journal} {\bibinfo  {journal} {Acta.
  Phys. Slovaca}\ }\textbf {\bibinfo {volume} {58}},\ \bibinfo {pages} {675}
  (\bibinfo {year} {2008})}\BibitemShut {NoStop}%
\bibitem [{\citenamefont {Chen-to}(1993)}]{dgfbookchento}%
  \BibitemOpen
  \bibfield  {author} {\bibinfo {author} {\bibfnamefont {T.}~\bibnamefont
  {Chen-to}},\ }\href@noop {} {\emph {\bibinfo {title} {Dyadic Green's
  functions in electromagnetic theory}}}\ (\bibinfo  {publisher} {IEEE PRESS
  Series on Electromagnetic Waves},\ \bibinfo {year} {1993})\BibitemShut
  {NoStop}%
\bibitem [{\citenamefont {Laliotis}\ \emph {et~al.}(2021)\citenamefont
  {Laliotis}, \citenamefont {Lu}, \citenamefont {Ducloy},\ and\ \citenamefont
  {Wilkowski}}]{laliotis2021review}%
  \BibitemOpen
  \bibfield  {author} {\bibinfo {author} {\bibfnamefont {A.}~\bibnamefont
  {Laliotis}}, \bibinfo {author} {\bibfnamefont {B.-S.}\ \bibnamefont {Lu}},
  \bibinfo {author} {\bibfnamefont {M.}~\bibnamefont {Ducloy}}, \ and\ \bibinfo
  {author} {\bibfnamefont {D.}~\bibnamefont {Wilkowski}},\ }\href@noop {}
  {\bibfield  {journal} {\bibinfo  {journal} {AVS Quantum Sci.}\ }\textbf
  {\bibinfo {volume} {3}},\ \bibinfo {pages} {043501} (\bibinfo {year}
  {2021})}\BibitemShut {NoStop}%
\bibitem [{sm()}]{sm}%
  \BibitemOpen
  \href@noop {} {}\bibinfo {note} {Details on the calculation of the
  quadrupole-quadrupole interaction in the retarded limit are given in the
  Supplemental Material [link to be inserted]}\BibitemShut {NoStop}%
\end{thebibliography}%

\end{document}


\title{Supplementary material to \\Multipole expansion for dispersion forces -- watch this trace}
\author{Jivesh Kaushal}
\affiliation{Institut für Physik, Universität Rostock, Albert-Einstein-Straße 23-24, D-18059 Rostock, 
Germany}
\author{Chris Boldt}
\affiliation{Institut für Physik, Universität Rostock, Albert-Einstein-Straße 23-24, D-18059 Rostock, 
Germany}
\author{Stefan Scheel}
\email{stefan.scheel@uni-rostock.de}
\affiliation{Institut für Physik, Universität Rostock, Albert-Einstein-Straße 23-24, D-18059 Rostock, 
Germany}
\author{Athanasios Laliotis}
\affiliation{Laboratoire de Physique des Lasers, Universit{\'e} Sorbonne Paris Nord, F-93430, Villetaneuse, France}
\affiliation{CNRS, UMR 7538, LPL, 99 Avenue J.-B. Cl{\'e}ment, F-93430 Villetaneuse, France} 
\author{Paolo Pedri}
\email{paolo.pedri@univ-paris13.fr}
\affiliation{Laboratoire de Physique des Lasers, Universit{\'e} Sorbonne Paris Nord, F-93430, Villetaneuse, France}
\affiliation{CNRS, UMR 7538, LPL, 99 Avenue J.-B. Cl{\'e}ment, F-93430 Villetaneuse, France} 

\maketitle

\section{Derivation of Eq.~(18)}

We start from the interaction Hamiltonian Eq.~(4):
\begin{equation}
\hat{H}_{\text{int}} = -\hat{\bd}\cdot\hat{\bE}(\br_A) - \hat{\bQ}\bullet\left[\nabla\otimes\hat{\bE}(\br_A)\right] - \hat{\bO}\bullet\left[\nabla\otimes\nabla\otimes\hat{\bE}(\br_A)\right].
\end{equation}
%
Following Sec.~5.2.1 in Scheel \cite{Acta} or Sec.~4.2 in Buhmann \cite{Buhmann}, we expand the interaction Hamiltonian for the octupole-dipole interaction over intermediate states $\vert I\rangle \equiv \vert k\rangle\vert \mathbf{1}_{\lambda}(\br,\omega)\rangle$, to obtain the perturbation energy arising from the octupole-dipole interaction:
\begin{equation}
\Delta E_{od} = \sum_{I\neq G}\frac{\langle G \vert \hat{H}_{\text{int}O}\vert I \rangle\langle I \vert \hat{H}_{\text{int}d} \vert G \rangle}{E_G - E_I}.
\end{equation}

We then use field expansions for the electric field operator $\hat{\bE}(\br_A)$ in terms of the scattering dyadic Green's function $\DGF_{\lambda}^{(1)}$ and the independent bosonic variable operators $\bff_{\lambda}$ \cite{Acta, Buhmann},
\begin{equation}
\hat{\bE}(\br_A) = \sum_{\lambda=e,m}\int_{}^{}d^3r' \DGF_{\lambda}^{(1)}(\br_A,\br',\omega)\cdot\bff_{\lambda}(\br',\omega),
\end{equation}
and from the properties of the bosonic operators $\bff$ \cite{Acta,Buhmann}, we get:
\begin{align}
\langle G \vert \hat{\bO}\bullet\nabla\otimes\nabla\otimes \hat{\bE}(\br_A) \vert I \rangle &= \langle 0 \vert\langle \{0\}\vert \hat{\bO}\bullet\nabla\otimes\nabla\otimes \hat{\bE}(\br_A) \vert \mathbf{1}_{\lambda}(\br,\omega)\rangle\vert k\rangle,\\
&=\bO_{0k}\bullet\nabla\otimes\nabla\otimes \DGF_{\lambda}^{(1)}(\br_A,\br,\omega).\\
\langle I \vert \hat{\bd}\cdot\hat{\bE}(\br_A) \vert G \rangle &= \langle k\vert\langle\mathbf{1}_{\lambda}(\br,\omega)\vert \hat{\bd}\cdot \hat{\bE}(\br_A) \vert \{0\}\rangle\vert 0\rangle,\\
&=\bd_{k0}\cdot \DGF_{\lambda}^{(1)}(\br_A,\br,\omega).
\end{align}
Then, following the steps in Sec. 5.2.1 in \cite{Acta}, we get:
\begin{equation}
U_{od} = -\frac{\mu_0}{\pi}\sum_{k}\int_{0}^{\infty}\frac{d\omega}{\omega_{k0}+\omega}\left[\omega^2 \bO_{0k}\bullet\nabla\otimes\nabla\otimes \DGF(\br_A,\br_A,\omega)\cdot\bd_{k0}\right].
\end{equation}
Taking advantage of expression of a general octupole moment in traceless form [Eq.~(12) in the main manuscript], and Eq.~(13) from the main manuscript, we finally arrive at Eq.~(18) in the main manuscript:
\begin{equation}
U_{od}^{\text{trace}}(\br_A) = \frac{\mu_0}{5\hbar c^2}\sum_{k}\int_{0}^{\infty}\frac{d\omega}{\omega_{k}+\omega}\omega^4 \bT_{0k}\cdot\Imm\DGF(\br_A,\br_A,\omega)\cdot\bd_{k0}.
\end{equation}

Finally, the $\omega$-integral is then rotated from the real frequency axis to the imaginary frequency axis ($\omega=i\xi$), to obtain the result
\begin{equation}
U_{od}^{\text{trace}}(\br_A) = \frac{\hbar\mu_0}{10\pi c^2}\int_{0}^{\infty}d\xi\,\xi^4\Tr[\bm{\alpha}_{od}^{(4)}(i\xi)\cdot\DGF(\br_A,\br_A,i\xi)],
\end{equation}
where the polarisability for the octupole-dipole moments is defined as
\begin{equation}
        \bm{\alpha}_{od}^{(4)}(i\xi) = \frac{1}{\hbar} \sum\limits_k \left( 
        \frac{\mathbf{d}_{k0}\otimes\mathbf{T}_{0k}}{i\xi+\omega_k} -
        \frac{\mathbf{T}_{0k}\otimes\mathbf{d}_{k0}}{i\xi-\omega_k} \right).
\end{equation}

Following the same steps as above, we can also derive the expression for dipole-octupole interaction potential,
\begin{equation}
U_{do}^{\text{trace}}(\br_A) = \frac{\hbar\mu_0}{10\pi c^2}\int_{0}^{\infty}d\xi\,\xi^4\Tr[\bm{\alpha}_{do}^{(4)}(i\xi)\cdot\DGF(\br_A,\br_A,i\xi)],
\end{equation}
with the definition of the dipole-octupole polarisability
\begin{equation}
\bm{\alpha}_{do}^{(4)}(i\xi) = \frac{1}{\hbar} \sum\limits_k \left( \frac{\mathbf{T}_{k0}\otimes\mathbf{d}_{0k}}{i\xi+\omega_k} -
\frac{\mathbf{d}_{0k}\otimes\mathbf{T}_{k0}}{i\xi-\omega_k} \right).
\end{equation}


\section{Quadrupole-Quadrupole retarded limit Eq.~(28)}
Here we give details on the computation of the quadrupole-quadrupole interaction potential
in the retarded limit that is used in the comparison with the trace part of the 
dipole-octupole potential. For that, we recall the ground-state Casimir-Polder potential
for quadrupole-quadrupole interactions \cite{Crosse}
\begin{equation}
\label{eq:cp}
    U_{qq}(z) = \frac{\hbar\mu_0}{2\pi} \int\limits_0^\infty d\xi\,\xi^2 \left[ 
    \bm{\alpha}_{qq}^{(4)}(i\xi) \bullet \bm{\nabla}\otimes 
    \bm{G}^{(1)}(\mathbf{r}_A,\mathbf{r}_A,i\xi) \otimes \overleftarrow{\bm{\nabla}}\right]
\end{equation}
where the integration extends over the positive imaginary frequency axis, and the symbol
$\bullet$ denotes the Frobenius product that contracts the quadrupole polarizability tensor
$\bm{\alpha}^{(4)}(i\xi)$ with the two-sided gradient of the scattering Green function
$\bm{G}^{(1)}(\mathbf{r}_A,\mathbf{r}_A,i\xi)$. In order to compute the gradients, we make
use of the Weyl expansion \cite{Acta}
\begin{equation}
\label{eq:weyl}
    \bm{G}^{(1)}(\mathbf{r},\mathbf{r}',i\xi) = \int d^2q \, 
    e^{i\mathbf{q}\cdot(\mathbf{r}-\mathbf{r}')} \bm{G}(\mathbf{q},z,z',i\xi) \,,\quad
    \mathbf{q}\perp\mathbf{e}_z\,,\quad \mathbf{e}_q=\cos\varphi \mathbf{e}_x +\sin\varphi 
    \mathbf{e}_y
\end{equation}
with the Fourier components
\begin{equation}
    \bm{G}(\mathbf{q},z,z',i\xi) = \frac{1}{8\pi^2b} \sum\limits_{\sigma=s,p}
    \mathbf{e}_\sigma^+ \otimes \mathbf{e}_\sigma^- r_\sigma\, e^{-b(z+z')} \,,\quad
    b^2 = q^2+ \frac{\xi^2}{c^2}
\end{equation}
with the sum running over the two transverse polarizations. The outer products of the
respective polarization unit vectors read
\begin{equation}
    \mathbf{e}_s^+ \otimes \mathbf{e}_s^- = 
    \begin{pmatrix} 
    \sin^2\varphi & -\sin\varphi \cos\varphi & 0 \\
    -\sin\varphi \cos\varphi & \cos^2\varphi & 0 \\
    0 & 0 & 1 
    \end{pmatrix}
    \,,\quad
    \mathbf{e}_p^+ \otimes \mathbf{e}_p^- = \frac{c^2}{\xi^2} 
    \begin{pmatrix} 
    b^2\cos^2\varphi & b^2\sin\varphi \cos\varphi & -ibq \cos\varphi \\
    b^2\sin\varphi \cos\varphi & b^2\sin^2\varphi & -ibq \sin\varphi \\
    ibq \cos\varphi & ibq \sin\varphi & q^2 
    \end{pmatrix} \,.
\end{equation}
Differentiating the Green function with respect to its spatial arguments leads to
additional factors under the $\mathbf{q}$-integral in Eq.~(\ref{eq:weyl}). Those factors are
\begin{eqnarray}
    \partial_x \partial_{x'} &\mapsto& q^2 \cos^2\varphi \,,\nonumber\\
    \partial_x \partial_{y'} &\mapsto& q^2 \sin\varphi \cos\varphi\,,\nonumber\\
    \partial_y \partial_{x'} &\mapsto& q^2 \sin\varphi \cos\varphi\,,\nonumber\\
    \partial_y \partial_{y'} &\mapsto& q^2 \sin^2\varphi \,,\nonumber\\
    \partial_x \partial_{z'} &\mapsto& -ibq \cos\varphi\,,\nonumber\\
    \partial_z \partial_{x'} &\mapsto& ibq \cos\varphi\,,\nonumber\\
    \partial_y \partial_{z'} &\mapsto& -ibq \sin\varphi\,,\nonumber\\
    \partial_z \partial_{y'} &\mapsto& ibq \sin\varphi\,,\nonumber\\
    \partial_z \partial_{z'} &\mapsto& b^2 \,.
\end{eqnarray}
The $q$-integral in Eq.~(\ref{eq:weyl}) is then converted into polar coordinates, with 
its angular part yielding elementary integral over powers of trigonometric functions of $\varphi$,
once the limit $\mathbf{r}'\to\mathbf{r}$ has been taken. The remaining integral over $q$ can be
converted using the substitution \cite{Safari}
\begin{equation}
    \int\limits_0^\infty dq \frac{q}{b} \ldots \mapsto \int\limits_1^\infty dv \frac{\xi}{c} \ldots
    \,,\quad b=v\frac{\xi}{c} \,,\quad q^2 = (v^2-1) \frac{\xi^2}{c^2} \,.
\end{equation}

At this point, we turn to the retarded limit in which the atom-surface distance is larger than 
any relevant interatomic transition frequencies.  In this limit, the polarizability takes on its
static value $\bm{\alpha}^{(4)}(0)$, and we also assume for simplicity the interface to be a 
perfect mirror with $r_s=-1$ and $r_p=+1$. Then, the frequency integral in Eq.~(\ref{eq:cp}) can 
be performed to give
\begin{equation}
    \int\limits_0^\infty d\xi \,\xi^5 e^{-2vz\xi/c} = \frac{15}{8} \left( \frac{c}{vz} \right)^6 \,,
\end{equation}
showing the well-known $z^{-6}$-dependence of the retarded Casimir-Polder potential for 
quadrupole-quadrupole interactions.

Before continuing to compute all derivatives of the Green function needed for the full 
Casimir-Polder potential (\ref{eq:cp}), we make another simplification by assuming that the 
polarizability can be taken as spatially isotropic. In this case, we perform a rotational
averaging based on the tensor contraction \cite{Andrews}
\begin{equation}
    \Bar{d}_{ij} \Bar{d}_{km} = I_{ijkm}^{\alpha\beta\gamma\rho} d_{\alpha\beta} d_{\gamma\rho}
    \,,\quad I_{ijkm}^{\alpha\beta\gamma\delta} = \frac{1}{30} 
    \begin{pmatrix}
        \delta_{ij} \delta_{km} \\ \delta_{ik} \delta_{jm} \\ \delta_{im} \delta_{jk}    
    \end{pmatrix}^T
    \begin{pmatrix}
        4 & -1 & -1 \\ -1 & 4 & -1 \\ -1 & -1 & 4
    \end{pmatrix}
    \begin{pmatrix}
        \delta_{\alpha\beta} \delta_{\gamma\rho} \\
        \delta_{\alpha\gamma} \delta_{\beta\rho} \\
        \delta_{\alpha\rho} \delta_{\beta\gamma}
    \end{pmatrix}\,.
\end{equation}
The quadrupole polarizability $\bm{\alpha}^{(4)}(0)$ contains exactly such combination of 
tensor products of quadrupole moments, hence we have to compute a quantity
\begin{equation}
    \Bar{d}_{ij} \Bar{d}_{km} \partial_i G_{jk} \partial_m = \frac{1}{30} 
    \begin{pmatrix}
        \partial_i G_{ij} \partial_j \\ \partial_i G_{ji} \partial_j \\ \partial_i G_{jj} \partial_i    
    \end{pmatrix}^T
    \begin{pmatrix}
        4 & -1 & -1 \\ -1 & 4 & -1 \\ -1 & -1 & 4
    \end{pmatrix}
    \begin{pmatrix}
        d_{\alpha\alpha} d_{\beta\beta} \\
        d_{\alpha\beta} d_{\alpha\beta} \\
        d_{\alpha\beta} d_{\beta\alpha}
    \end{pmatrix}\,.
\end{equation}
Now we use the fact that the quadrupole moments are traceless, hence $d_{\alpha\alpha}=0$, and that
the scattering Green function is double-sided transverse, hence 
$\partial_i G_{ij} \partial_j=\partial_i G_{ji} \partial_j=0$. This leaves us with
\begin{equation}
    \Bar{d}_{ij} \Bar{d}_{km} \partial_i G_{jk} \partial_m = \frac{1}{30} 
    \begin{pmatrix}
        0 \\ 0 \\ \partial_i G_{jj} \partial_i    
    \end{pmatrix}^T
    \begin{pmatrix}
        4 & -1 & -1 \\ -1 & 4 & -1 \\ -1 & -1 & 4
    \end{pmatrix}
    \begin{pmatrix}
        0 \\
        d_{\alpha\beta} d_{\alpha\beta} \\
        d_{\alpha\beta} d_{\beta\alpha}
    \end{pmatrix}
    = \frac{1}{10} d_{\alpha\beta} d_{\alpha\beta} \partial_i G_{jj} \partial_i\,,
\end{equation}
where we have used the symmetry of the quadrupole moments. Hence, it suffices to compute the 
double-sided derivatives of the trace of the scattering Green function. The result of this rather
tedious but straightforward calculation is the Casimir-Polder potential for quadrupole-quadrupole
interactions in the retarded limit as
\begin{equation}
    U_{qq}(z) = -\frac{5\hbar c}{64\pi^2\varepsilon_0 z^6} \alpha_{qq}(0) \,.
\end{equation}

Similarly, the isotropic average of the octupole-dipole trace term yields \cite{Crosse}
\begin{equation}
U_{od}(z) = -\frac{\hbar c}{96\pi^2\varepsilon_0 z^6} \alpha_{od}(0),\,\text{and}\quad U_{do}(z) = -\frac{\hbar c}{96\pi^2\varepsilon_0 z^6} \alpha_{do}(0),\
\end{equation}
where the averaging here is $\Bar{d}_i\Bar{d}_j=\frac{1}{3}\delta_{ij}d_\alpha d_\alpha$.
Therefore, the ratio $(U_{od} + U_{do})/U_{qq}$ is given by
\begin{equation}
\frac{U_{od}(z) + U_{do}(z)}{U_{qq}(z)} = \frac{2}{15}\frac{\alpha_{od}(0)+ \alpha_{do}(0)}{\alpha_{qq}(0)}. \label{eqn:ratio}
\end{equation}

Using the data from the Alkali-Rydberg-Calculator \cite{ARC}, we sum over the nearest principal quantum numbers $n$ around the atomic ground state $\vert 6S_{1/2}\rangle$ of Cesium. For the quadrupole-quadrupole transitions, we considered all possible substates from $\vert 5D_J m_J\rangle$ until $\vert 9D_J m_J\rangle$ ($m_J \in [-J, J]\,\forall J \in \{3/2,5/2\}$). For the octupole-dipole transitions, we considered all possible substates from $\vert 6P_J m_J\rangle$ until $\vert 10P_J m_J\rangle$ ($m_J \in [-J, J]\,\forall J\in \{5/2,7/2\}$). This yields for the ratio of the polarizabilities
\begin{equation}
\frac{\alpha_{od}(0)+ \alpha_{do}(0)}{\alpha_{qq}(0)} = 3.915.  
\end{equation}
Combined with Eq.~\eqref{eqn:ratio}, we obtain the ratio of the respective interaction potentials as
\begin{equation}
\frac{U_{od}(z) + U_{do}(z)}{U_{qq}(z)} = \frac{2}{15}\times 3.915 \approx 0.52.
\end{equation}